\documentclass[preprint,12pt]{elsarticle}

\usepackage{amssymb}
\usepackage{amsmath}
\usepackage{appendix}

\begin{document}
\begin{frontmatter}

\title{Quantum Public-Key Encryption with Information Theoretic Security}
\author{Jiangyou Pan}
\author{Li Yang\corref{1}}\ead{yangli@iie.ac.cn}
\cortext[1]{Corresponding author.}
\address[]{State Key Laboratory of Information Security, Graduate University of Chinese Academy of Sciences,
Beijing 100049, China}

\begin{abstract}
We propose a definition for the information theoretic security of a quantum public-key encryption scheme, and present bit-oriented and two-bit-oriented encryption schemes satisfying our security definition via the introduction of a new public-key algorithm structure. We extend the scheme to a multi-bit-oriented one, and conjecture that it is also information theoretically secure, depending directly on the structure of our new algorithm.
\end{abstract}


\end{frontmatter}

\makeatletter
    \newcommand{\rmnum}[1]{\romannumeral #1}
    \newcommand{\Rmnum}[1]{\expandafter\@slowromancap\romannumeral #1@}
    \newcommand{\bm}[1]{\mbox{\boldmath{$#1$}}}
\makeatother

\section{Introduction}
\noindent
The public-key encryption schemes currently used will not keep their security in the post-quantum era, so it is necessary to find new kinds of encryption to resist the attacks of quantum adversaries. Quantum public-key encryption (QPKE) is one solution, which has been studied for about ten years. Okamoto et al \cite{1} put forward a knapsack-based scheme which involves a quantum algorithm during key generation. Gottesman and Chuang \cite{2} were the first to use quantum states as public keys. Gottesman was also the first to put forward "Quantum Public Key Cryptography with Information-Theoretic Security" \cite{3} based on  Einstein-Podolsky-Rosen  pairs. Yang \cite{4} has discussed public-key encryption of quantum messages based on a classical computational complexity hypothesis. Kawachi et al \cite{5} investigated the cryptographic property "computational indistinguishability" of two quantum states generated via fully flipped permutations ($QSCD_{ff}$), and gave a QPKE scheme  based on this. Nikolopoulos \cite{6} suggested another scheme based on qubit rotations. The latter two schemes are bit-oriented, and Kawachi et al extended their scheme to multibits \cite{7},which was later shown in \cite{8} to have bounded information theoretic security.

\section{Security of Quantum Public-Key Encryption}
\noindent
In classical public-key encryption (PKE), the ciphertext indistinguishability under a chosen plaintext attack (CPA) is defined as \cite{9}: for every polynomial-size circuit family $\{C_n\}$, every positive polynomial $p(\cdot)$, all sufficiently large $n$, and every $x,y$ in plaintext space,the probability $\Pr(\cdot)$ satisfies:
\begin{eqnarray}
|\Pr[C_n(G_1(1^n),E_{G_1(1^n)}(x))=1]-\Pr[C_n(G_1(1^n),E_{G_1(1^n)}(y))=1]|<\frac{1}{p(n)}.
\end{eqnarray}

As the ciphertext is a quantum state in the quantum case, the ciphertext indistinguishability of QPKE is defined as the indistinguishability of any two quantum states in ciphertext space. Koshiba \cite{10} extended indistinguishability to QPKE, but he restricted the circuit to a polynomial-size one. We propose here a definition of ciphertext indistinguishability of quantum public-key encryption beyond the computational complexity hypothesis.

\vspace*{12pt}
\newtheorem{defn}{Definition}
\begin{defn}
A quantum public-key encryption has ciphertext indistinguishability under CPA, if for every quantum circuit family $\{C_n\}$, for every positive polynomial $p(\cdot)$, all sufficiently large $n$, and every $x,y$ in plaintext space,the probability $\mathrm{Pr}(\cdot)$ satisfies:
\begin{eqnarray}
|\Pr[C_n(G_1(1^n),E_{G_1(1^n)}(x))=1]-\Pr[C_n(G_1(1^n),E_{G_1(1^n)}(y))=1]|<\frac{1}{p(n)}.
\label{eqn:one}
\end{eqnarray}
where the encryption algorithm $E$ is a quantum algorithm, and the ciphertexts $E(x)$ and $E(y)$ are quantum states.
\end{defn}
\vspace*{12pt}

The difference between our definition and Koshiba's is that there is no restriction on $\{C_n\}$ in our definition.

According to \cite{9}(see page 476), the definition we have presented here is related to information theoretic security under CPA. We define: A quantum public-key encryption is information theoretically secure under CPA if it satisfies Eq. (\ref{eqn:one}).

In the following part, we give a quantum public-key encryption scheme which satisfies our definition of information theoretic security under CPA.

\section{A bit oriented Public-key Encryption Scheme}
\noindent
Let:
$\Omega_n=\{k\in Z_{2^n}|\ W_H(k)\ is\ odd\}$ and $\Pi_n=\{k\in Z_{2^n}|\ W_H(k)\ is\ even\}$, where $W_H(k)$ is $k$'s Hamming weight.

\vspace*{12pt}
\begin{defn}
Define two $n$-qubit states:
\begin{equation}
\rho_{k,i}^0=\frac{1}{2}(|i\rangle+|i\oplus k\rangle)(\langle i|+\langle i\oplus k|),
\end{equation}
and
\begin{equation}
\rho_{k,i}^1=\frac{1}{2}(|i\rangle-|i\oplus k\rangle)(\langle i|-\langle i\oplus k|),
\end{equation}
where $i\in Z_{2^n}$,$k\in\Omega_n$ .
\end{defn}
\vspace*{12pt}

The two states $\rho_{k,i}^0$ and $\rho_{k,i}^1$ can be generated effectively as follows:

For given $i$ and $k$, use a permutation operator $P_k$ on $|k\rangle$,
so that $P_k|k\rangle=|1\cdot\cdot\cdot 10\cdot\cdot\cdot 0\rangle$. Let $P_k|i\rangle =|i'\rangle |i''\rangle$, so:

\begin{eqnarray}
\frac{1}{\sqrt{2}}P_k(|i\rangle+|i\oplus k\rangle)
&=&\frac{1}{\sqrt{2}}(|i'\rangle |i''\rangle+|i'\oplus(2^{W_H(k)}-1)\rangle|i''\oplus0\rangle) \nonumber\\
&=&\frac{1}{\sqrt{2}}(|i'\rangle+|\overline{i'}\rangle)|i''\rangle,
\end{eqnarray}
where $|\overline{i'}\rangle$ is the state after applying the $X$ operation on each qubit of $|i'\rangle$.

It can be seen that, $\frac{1}{\sqrt{2}}(|i'\rangle+|\overline{i'}\rangle)$ is a $ W_H(k)$ bits GHZ state. Inverting the above process, we obtain an effective way to generate $\rho_{k,i}^0$ if GHZ states are available.

To produce $\rho_{k,i}^1$, we just apply $Z$ on each qubit of $\rho_{k,i}^0$ for $W_H(k)$ odd. Then we have a polynomial-time quantum algorithm to convert $\rho_{k,i}^0$ to $\rho_{k,i}^1$ without $k$ and $i$.

\subsection{Application in Quantum Public-Key Encryption}
\noindent
Our quantum public-key encryption is shown in  Figure \ref{fig1}:

\begin{figure}[htp!]
\begin{center}
\includegraphics[width=8.2cm]{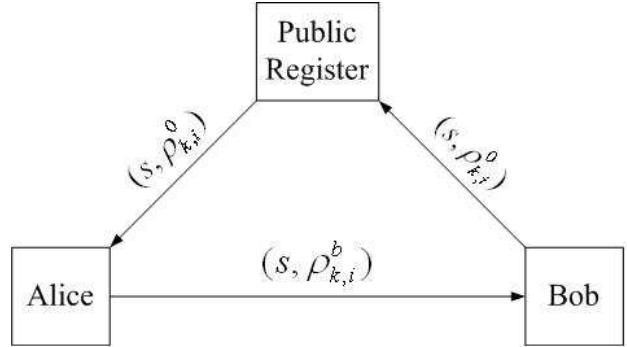}
\end{center}
\caption{\label{fig1}First, Bob sends his public-key $(s,\rho_{k,i}^0)$ to a public register. Alice gets Bob's public-key from the public register, then she encrypts $b$ into $\rho_{k,i}^b$, and sends $(s,\rho_{k,i}^b)$ back to Bob.}
\end{figure}

\begin{bfseries}
\flushleft{[Key Generation]}
\end{bfseries}

\begin{enumerate}
\renewcommand{\labelenumi}{(G\theenumi)}
\item Bob selects randomly a Boolean function $F:\Omega_n\to\Omega_n$ as private key;
\item Bob selects $s\in\Omega_n$ randomly;
\item Bob generates $\rho_{k,i}^0$, where $k=F(s)$, $i$ is chosen randomly from $Z_{2^n}$;
\item Bob sends the classical-quantum pair $(s,\rho_{k,i}^0)$ to a public register as his public-key.
\end{enumerate}

When Alice needs to send a classical bit $b$ to Bob via the quantum channel, they can do as follows:

\begin{bfseries}
\flushleft{[Encryption]}
\end{bfseries}

\begin{enumerate}
\renewcommand{\labelenumi}{(E\theenumi)}
\item Alice gets one of Bob's public keys from the public register;
\item Alice encrypts $b$ into $\rho_{k,i}^b$, then sends $(s,\rho_{k,i}^b)$ to Bob;
\end{enumerate}

\begin{bfseries}
\flushleft{[Decryption]}
\end{bfseries}

\begin{enumerate}
\renewcommand{\labelenumi}{(D\theenumi)}
\item Bob receives $(s,\rho_{k,i}^b)$, and calculates $k=F(s)$;
\item Bob decrypts the one-bit message with $k$: $(s,\rho_{k,i}^b)\to b$.
\end{enumerate}
Notes:
\begin{enumerate}
\renewcommand{\labelenumi}{(\theenumi)}
\item The Boolean function $F$ can be chosen from a larger set $\{0,1\}^{poly(n)}$, but when $s$ is chosen, it should satisfy $F(s)\in\Omega_n$.
\item The public register should ensure that Alice obtains the correct public-key from Bob. This is a precondition of all public-key encryption schemes.
\end{enumerate}

\subsection{Trapdoor Property}
\noindent
Bob can decrypt the cipertext states with $k$, but without $i$. So, we consider the mixed states: $\rho_k^0=\frac{1}{2^n}\sum_i \rho_{k,i}^0$ and $\rho_k^1=\frac{1}{2^n}\sum_i \rho_{k,i}^1$.
\vspace*{12pt}
\newtheorem{lem}[defn]{Lemma}
\begin{lem}
\label{thm:two}
For given $k\in \Omega_n$, there exists a polynomial-time quantum algorithm that distinguishes   $\rho_k^0$ and $\rho_k^1$ determinedly.
\end{lem}
\vspace*{12pt}
\noindent
{\bf Proof:}
Let $\rho$ be the unknown state $\rho_k^0$ or $\rho_k^1$, then the algorithm is given as follows:
\begin{enumerate}
\renewcommand{\labelenumi}{(\theenumi)}
\item  Prepare two quantum registers, the first register holds a control bit in $|0\rangle \langle 0|$, and the second one holds $\rho$. After Hadamard's transformation is applied to the first register, the state of the system becomes:
    \begin{equation}
    \frac{(|0\rangle+|1\rangle)(\langle 0|+\langle 1|)}{2}\otimes\rho.
    \end{equation}

\item Define Controlled-$k$ operator $C_k$ as: $C_k|0\rangle|i\rangle=|0\rangle|i\rangle$, $C_k|1\rangle|i\rangle=|1\rangle|i\oplus k\rangle$
    for any $i\in Z_{2^n}$, ($C_k$ can be realized via a group of CNOT operations), apply $C_k$ to the two registers, then the result will be
    \begin{equation}
    \frac{1}{2^n}\sum_{i=0}^{2^n-1}|\varphi_{k,i}^0\rangle\langle\varphi_{k,i}^0|,\ if \ \rho=\rho_k^0,
    \end{equation} or
    \begin{equation}
    \frac{1}{2^n}\sum_{i=0}^{2^n-1}|\varphi_{k,i}^1\rangle\langle\varphi_{k,i}^1|,\ if \ \rho=\rho_k^1,
    \end{equation}
    where
    \begin{eqnarray}
    |\varphi_{k,i}^0\rangle &=& C_k[\frac{1}{2}(|0\rangle+|1\rangle)(|i\rangle+|i\oplus k\rangle)] \nonumber\\
    &=&\frac{1}{2}\left[|0\rangle(|i\rangle +|i\oplus k\rangle)+|1\rangle(|i\oplus k\rangle+|i\rangle)\right],
    \end{eqnarray}

    \begin{eqnarray}
    |\varphi_{k,i}^1\rangle &=& C_k[\frac{1}{2}(|0\rangle+|1\rangle)(|i\rangle-|i\oplus k\rangle)] \nonumber\\
    &=&\frac{1}{2}\left[|0\rangle(|i\rangle -|i\oplus k\rangle)+|1\rangle(|i\oplus k\rangle-|i\rangle)\right].
    \end{eqnarray}

\item Apply Hadamard transformation to the first register again:
    \begin{eqnarray}
    (H\otimes I)|\varphi_{k,i}^0\rangle=\frac{1}{\sqrt{2}}|0\rangle(|i\rangle+|i\oplus k\rangle),
    \end{eqnarray}
    \begin{eqnarray}
    (H\otimes I)|\varphi_{k,i}^1\rangle=\frac{1}{\sqrt{2}}|1\rangle(|i\rangle-|i\oplus k\rangle).
    \end{eqnarray}
    If $\rho=\rho_k^0$, the final state is $|0\rangle\langle0|\otimes\rho_k^0$; if $\rho=\rho_k^1$, the final state is $|1\rangle\langle1|\otimes\rho_k^1$.
\end{enumerate}

It can be seen that we can distinguish $\rho_k^0$and $\rho_k^1$  with correct probability 1 after measuring the first register. $\hfill{}\Box$

\subsection{Security Proof}
\noindent
If there exists an eavesdropper Eve between Alice and Bob, she may use two ways to attack the QPKE. One is to find information about $k$; another is to distinguish between $\rho_{k,i}^0$ and $\rho_{k,i}^1$ to eavesdrop the message.

By measuring $\rho_{k,i}^0$ or $\rho_{k,i}^1$, Eve can get $i$ or $i\oplus k$ with the same probability $1/2$, but she cannot get both of them. For each $\rho_{k,i}^0$ or $\rho_{k,i}^1$, $i$ and $s$ are chosen randomly, then $k=F(s)$ is also random. Although Eve may get $i\oplus k$ with probability $1/2$, she cannot obtain any information about $k$ because she cannot get $i$ at the same time. The security is the same as that of a one-time-pad in classical cryptography.

If Eve has an effective algorithm to distinguish $\rho_{k,i}^0$ and $\rho_{k,i}^1$ with non-negligible probability without $k$ and $i$, that means Eve can distinguish the mixed states:  $\rho_{odd}^0=\frac{1}{2^{n-1}\cdot2^n}\sum_{k\in \Omega_n}\sum_i \rho_{k,i}^0$, and
$\rho_{odd}^1=\frac{1}{2^{n-1}\cdot2^n}\sum_{k\in \Omega_n}\sum_i \rho_{k,i}^1$. However, we have the following lemma:
\vspace*{12pt}
\begin{lem}
The trace distance (defined as \cite{11}) between $\rho_{odd}^0$ and $\rho_{odd}^1$ is $\frac{1}{2^{n-1}}$.
\label{thm:one}
\end{lem}
\vspace*{12pt}
\noindent
{\bf Proof:}
\begin{equation}
\rho_{odd}^0-\rho_{odd}^1
=\frac{4}{2^{2n}}\sum_{k\in \Omega_n}\sum_i|i\rangle\langle i\oplus k|
=\frac{4}{2^{2n}}A_{odd},
\end{equation}
where $A_{odd}$ is a matrix with
\begin{equation}
a_{ij}=\big\{{{1,\ \ W_H(i)\ mod\ 2\ne W_H(j)\ mod\ 2} \atop {0,\ \ W_H(i)\ mod\ 2= W_H(j)\ mod\ 2}}.
\label{eqn:Aodd}
\end{equation}
Applying an appropriate unitary operator to both sides of $A_{odd}$, we obtain
\begin{equation}
A'=\left[\begin{array}{ccccc}
0&1&\cdots&0&1\\
1&0&\cdots&1&0\\
&&\cdots&&\\
0&1&\cdots&0&1\\
1&0&\cdots&1&0
\end{array}\right]
=\left[\begin{array}{cc}
1&1\\
1&1
\end{array}\right]^{\otimes{(n-1)}}
\otimes\left[\begin{array}{cc}
0&1\\
1&0
\end{array}\right].
\end{equation}
For $|A\otimes B|=|A|\otimes|B|$ and $tr(A\otimes B)=tr(A)\times tr(B)$, we have:
\begin{eqnarray}
tr|A_{odd}|=tr|A'|
=tr\left|\left[\begin{array}{cc}
1&1\\
1&1
\end{array}\right]\right|^{(n-1)}
\times tr\left|\left[\begin{array}{cc}
0&1\\
1&0
\end{array}\right]\right|
=2^n,
\label{eqn:trAodd}
\end{eqnarray}
\begin{eqnarray}
D(\rho_{odd}^0,\rho_{odd}^1)=\frac{1}{2}tr\left|\frac{4}{2^{2n}}A_{odd}\right|
=\frac{4}{2\cdot 2^{2n}}\cdot 2^{n}
=\frac{1}{2^{n-1}}.
\end{eqnarray}
$\hfill{}\Box$

\vspace*{12pt}
\newtheorem{thm}[defn]{Theorem}
\begin{thm}
The quantum public-key encryption given above satisfies the inequality (\ref{eqn:one}), so it has information theoretic security under CPA.
\label{thm:three}
\end{thm}
\vspace*{12pt}
\noindent
{\bf Proof:}
For every quantum circuit family $\{C_n\}$, and every $x,y\in \{0,1\}$,
\begin{eqnarray}
\Pr[C_n(G_1(1^n),E_{G_1(1^n)}(x))=1]
&=&\Pr\limits_{k,i}[C_n(\rho_{k,i}^x)=1]\nonumber\\
&=&\sum_{k,i}p_{k,i}\cdot \Pr[C_n(\rho_{k,i}^x)=1]\nonumber\\
&=&\frac{1}{2^{2n-1}}\sum_{k,i}\Pr[C_n(\rho_{k,i}^x)=1]\nonumber\\
&=&\Pr[C_n(\frac{1}{2^{2n-1}}\sum_{k,i}\rho_{k,i}^x)=1]\nonumber\\
&=&\Pr[C_n(\rho_{odd}^x)=1].
\end{eqnarray}
If $x$ and $y$ are different, we consider the difference between $\rho_{odd}^0$ and $\rho_{odd}^1$.

Any quantum circuit family $\{C_n\}$ that distinguishes between quantum states $\rho_{odd}^0$ and $\rho_{odd}^1$ can be regarded as distinguishing two probability distributions $\{p_m\}$ and $\{q_m\}$ based on a positive operator-valued measure(POVM) $\{E_m\}$ \cite{12,13}, where $p_m=tr(C_n(\rho_{odd}^0)E_m)$ and $q_m=tr(C_n(\rho_{odd}^1)E_m)$ are the probability distributions of quantum measurement outcomes labeled by $m$. The maximum trace distance between $\{p_m\}$ and $\{q_m\}$ \cite{11} over the whole set of POVMs determines the probability upper bound for distinguishing $\rho_{odd}^0$ and $\rho_{odd}^1$ by $\{C_n\}$,
\begin{eqnarray}
&&|\Pr[C_n(\rho_{odd}^0)=1]-\Pr[C_n(\rho_{odd}^1)=1]| \nonumber\\
&\le& \max\limits_{\{E_m\}} \frac{1}{2}
\sum_m |tr[E_m(C_n(\rho_{odd}^0)-C_n(\rho_{odd}^1))]|\nonumber\\
&=& \max\limits_{\{E_m\}} D(p_m,q_m).
\end{eqnarray}
According to \cite{11,14},
\begin{eqnarray}
\max\limits_{\{E_m\}} D(p_m,q_m)=D(C_n(\rho_{odd}^0),C_n(\rho_{odd}^1))\le D(\rho_{odd}^0,\rho_{odd}^1)=\frac{1}{2^{n-1}}.
\end{eqnarray}
For any positive polynomial $p(\cdot)$, there exists a sufficiently large $n$ so that Eq. (\ref{eqn:one}) is satisfied.$\hfill{}\Box$

\textbf{Remark.}
Like the public-key encryption mentioned above, in \cite{5,7}, the private-key $\pi$ and public-key $\rho_\pi^+$ have a one-to-one correspondence, so the public-key $\rho_{\pi}^+$ can only be used $t$ times, $t=o(nlogn)$ \cite{8}. Because $\pi$ contains $O(nlogn)$-bits, its efficiency is no better than a one-time pad. In our scheme, the private-key $F$ is about $poly(n)$-bits long, and it corresponds to a group of public-keys $(s,\rho_{i,k}^0)$, where $s$ and $i$ are chosen randomly. As mentioned above, $s$ is open but $k=F(s)$ is hidden by the one-time key $i$, so the adversary cannot compute $F$. Our private-key can be reused $2^{O(n)}$ times.

If we change our scheme to one similar to that in \cite{5}, letting $k$ be the private-key, $\rho_k^0$ the public-key, then using the method in Theorems 2.4 and 3.1 of \cite{8}, we can obtain following result:
\vspace*{12pt}
\begin{thm}
If we fix the key pair at $(\rho_k^0,k)$, the key pair can only be used t times, t=o(n).
\end{thm}
\vspace*{12pt}
\noindent
{\bf Proof:}
We calculate $\|\frac{1}{2^{n-1}} \sum_k {(\rho_k^0-\rho_k^1)\otimes(\rho_k^0)^{\otimes t})}\|_{tr}$. For simplicity, we calculate $\|\frac{1}{2^{n-1}}\sum_k \rho_k^0\otimes(\rho_k^0)^{\otimes t}-(\frac{I}{2^n})^{\otimes t+1}\|_{tr}$ and use the triangle inequality.

\begin{eqnarray}
\rho_k^0=\frac{1}{2 \cdot 2^n} \sum_i 2(|i\rangle\langle i|+|i\rangle\langle i\oplus k|)
=\frac{1}{2^n} \sum_i \sum_x |i\rangle(\langle i\oplus xk|),
\end{eqnarray}
where $x\in\{0,1\}$. Thus we have
\begin{eqnarray}
&&\|\frac{1}{2^{n-1}} \sum_k \left((\rho_k^0)^{\otimes t}- (\frac{I}{2^n})^{\otimes t}\right)\|_{tr} \nonumber\\
&=&\frac{1}{2^{n-1}\cdot 2^{nt}} \| \sum_k \sum_{i_1,\cdots,i_t} \sum_{x_1,\cdots,x_t} \left(|i_1,\cdots,i_t\rangle \langle i_1\oplus x_1 k,\cdots,i_t\oplus x_t k|- \right.  \nonumber\\
&& \left. -|i_1,\cdots,i_t\rangle \langle i_1,\cdots,i_t|\right) \|_{tr} \nonumber\\
&=&\frac{1}{2^{n-1}\cdot 2^{nt}} \|\sum_k \sum_{i_1,\cdots,i_t} \sum_{{x_1,\cdots,x_t} \atop {(x_1,\cdots,x_t)\ne(0,\cdots,0)}} |i_1,\cdots,i_t\rangle \langle i_1\oplus x_1 k,\cdots,i_t\oplus x_t k|\|_{tr} \nonumber\\
&\le&\frac{1}{2^{n-1}\cdot 2^{nt}} \sum_{i_1,\cdots,i_t} \||i_1,\cdots,i_t\rangle\|\cdot
\|\sum_k \sum_{{x_1,\cdots,x_t} \atop {(x_1,\cdots,x_t)\ne(0,\cdots,0)}} \langle i_1\oplus x_1 k,\cdots,i_t\oplus x_t k|\| \nonumber\\
&=&\frac{1}{2^{n-1}\cdot 2^{nt}} \cdot 2^{nt} \cdot \sqrt{2^{n-1}(2^t-1)} \nonumber\\
&<&\sqrt{\frac {1}{2^{n-t}}}.
\end{eqnarray}
If we want $\|\frac{1}{2^{n-1}} \sum_k {(\rho_k^0-\rho_k^1)\otimes(\rho_k^0)^{\otimes t})}\|_{tr}<1/p(n)$, $t$ should be $o(n)$.$\hfill{}\Box$

Our QPKE is information theoretic security, which is realized via a new public-key algorithm structure. The security of the scheme in \cite{8} is bounded information theoretic secure because it is based on a common public-key structure.

\section{Extended QPKE for Multibits}
\noindent
We take a two-bit scheme as an example to show how to extend our QPKE to encrypt more than one bit with each pair of the public-key.

\subsection{Four States Used to Construct the Public-Key}
\noindent
\begin{defn}
\label{def3}
Define the $n$-qubit state as:
\begin{eqnarray}
|\Psi_{k_1,k_2,i}^{00}\rangle&=&\frac{1}{2}(|i\rangle+|i\oplus k_1\rangle+|i\oplus k_2\rangle + |i\oplus k_1\oplus k_2\rangle)\nonumber\\
&=&\frac{1}{2}(|i_1\rangle|i_2\rangle+|i_1\oplus k_{11}\rangle|i_2\oplus k_{12}\rangle
+|i_1\oplus k_{21}\rangle|i_2\oplus k_{22}\rangle\nonumber\\
&+&|i_1\oplus k_{11}\oplus k_{21}\rangle|i_2\oplus k_{12}\oplus k_{22}\rangle),
\end{eqnarray}
where $k_1,k_2,i \in Z_{2^n}$, $i_1,i_2\in Z_{2^{\frac{n}{2}}}$, $k_{11},k_{22}\in\Omega_{2^{\frac{n}{2}}}$, $k_{12},k_{21}\in\Pi_{2^{\frac{n}{2}}}$, and $i=(i_1,i_2)$, $k_1=(k_{11},k_{12})$, $k_2=(k_{21},k_{22})$.
\end{defn}
\vspace*{12pt}

Applying $I^{\otimes\frac{n}{2}}\otimes Z^{\otimes\frac{n}{2}}$ on $|\Psi_{k_1,k_2,i}^{00}\rangle$, we obtain:
\begin{equation}
|\Psi_{k_1,k_2,i}^{01}\rangle=\frac{1}{2}(|i\rangle+|i\oplus k_1\rangle-|i\oplus k_2\rangle - |i\oplus k_1\oplus k_2\rangle).
\end{equation}

Applying $Z^{\otimes\frac{n}{2}}\otimes I^{\otimes\frac{n}{2}}$ on  $|\Psi_{k_1,k_2,i}^{00}\rangle$, we obtain:
\begin{equation}
|\Psi_{k_1,k_2,i}^{10}\rangle=\frac{1}{2}(|i\rangle-|i\oplus k_1\rangle+|i\oplus k_2\rangle - |i\oplus k_1\oplus k_2\rangle).
\end{equation}

Applying  $Z^{\otimes n}$ on  $|\Psi_{k_1,k_2,i}^{00}\rangle$, we obtain:
\begin{equation}
|\Psi_{k_1,k_2,i}^{11}\rangle=\frac{1}{2}(|i\rangle-|i\oplus k_1\rangle-|i\oplus k_2\rangle + |i\oplus k_1\oplus k_2\rangle).
\end{equation}

Let the four states be the cipher text of two classical bits. We construct a two-bit oriented QPKE scheme based on them.

\subsection{Trapdoor Property}
\noindent
Suppose $k_1$ and $k_2$ are given, for which without $i$ the four mixed states are:
\begin{eqnarray}
\rho_{k_1,k_2}^{00}=\frac{1}{2^n}\sum_i |\Psi_{k_1,k_2,i}^{00}\rangle\langle\Psi_{k_1,k_2,i}^{00}|,
\end{eqnarray}
\begin{eqnarray}
\rho_{k_1,k_2}^{01}=\frac{1}{2^n}\sum_i |\Psi_{k_1,k_2,i}^{01}\rangle\langle\Psi_{k_1,k_2,i}^{01}|,
\end{eqnarray}
\begin{eqnarray}
\rho_{k_1,k_2}^{10}=\frac{1}{2^n}\sum_i |\Psi_{k_1,k_2,i}^{10}\rangle\langle\Psi_{k_1,k_2,i}^{10}|,
\end{eqnarray}
\begin{eqnarray}
\rho_{k_1,k_2}^{11}=\frac{1}{2^n}\sum_i |\Psi_{k_1,k_2,i}^{11}\rangle\langle\Psi_{k_1,k_2,i}^{11}|.
\end{eqnarray}

We take $\rho_{k_1,k_2}^{10}$ as an example to explain that the algorithm described in Lemma \ref{thm:two} can also be used to distinguish these four states. The process includes the following steps:
\begin{enumerate}
\renewcommand{\labelenumi}{(\theenumi)}
\item Prepare two quantum registers, the first one contains two control bits in state $|0\rangle\langle 0|\otimes|0\rangle \langle 0|$, and the second one contains the unknown state. Take $\rho_{k_1,k_2}^{10}$ as an example, then the state of the system is
    \begin{equation}
    |0\rangle \langle 0|\otimes|0\rangle \langle 0|\otimes\rho_{k_1,k_2}^{10}.
    \end{equation}

\item Apply the Hadamard transformation to the first control bit and the controlled-$k_1$ operator to $\rho_{k_1,k_2}^{10}$, then the state of the system becomes
    \begin{equation}
    \frac{1}{2^n}\sum_i |\varphi_{k_1,k_2,i}^{10}\rangle\langle\varphi_{k_1,k_2,i}^{10}|,
    \end{equation}
    where
    \begin{eqnarray}
    |\varphi_{k_1,k_2,i}^{10}\rangle
    =\frac{1}{2\sqrt{2}}&[|0\rangle|0\rangle(|i\rangle-|i\oplus k_1\rangle+|i\oplus k_2\rangle - |i\oplus k_1\oplus k_2\rangle)\nonumber\\
    &+|1\rangle|0\rangle(|i\oplus k_1\rangle-|i\rangle
    -|i\oplus k_1\oplus k_2\rangle+|i\oplus k_2\rangle)].
    \end{eqnarray}

\item Apply the Hadamard transformation to the first control bit again, and the state of the system becomes
    \begin{equation}
    |1\rangle \langle 1|\otimes|0\rangle \langle 0|\otimes\rho_{k_1,k_2}^{10}.
    \end{equation}

\item It can be seen that if we perform operations $H\otimes I\cdot C_{k_2}\cdot H\otimes I$ that is related to the second control bit, the final state of the system will be
    \begin{equation}
    |1\rangle \langle 1|\otimes|0\rangle \langle 0|\otimes\rho_{k_1,k_2}^{10}.
    \end{equation}
\end{enumerate}

If the unknown state is one of the other three, the final state will be:
    \begin{equation}
    |0\rangle \langle 0|\otimes|0\rangle \langle 0|\otimes\rho_{k_1,k_2}^{00},
    \end{equation}

    or
    \begin{equation}
    |0\rangle \langle 0|\otimes|1\rangle \langle 1|\otimes\rho_{k_1,k_2}^{01},
    \end{equation}

    or
    \begin{equation}
    |1\rangle \langle 1|\otimes|1\rangle \langle 1|\otimes\rho_{k_1,k_2}^{11}.
    \end{equation}

We can distinguish between these four states by measuring the first register.

\subsection{Indistinguishability Property}
\noindent
Without $k_1,k_2$ and $i$, the ciphertext consists of four mixed states:
\begin{equation}
\rho_{odd}^{00}=\frac{1}{2^{2n-4}}\sum_{k_1,k_2} \rho_{k_1,k_2,i}^{00},
\end{equation}
\begin{equation}
\rho_{odd}^{01}=\frac{1}{2^{2n-4}}\sum_{k_1,k_2} \rho_{k_1,k_2,i}^{01},
\end{equation}
\begin{equation}
\rho_{odd}^{10}=\frac{1}{2^{2n-4}}\sum_{k_1,k_2} \rho_{k_1,k_2,i}^{10},
\end{equation}
\begin{equation}
\rho_{odd}^{11}=\frac{1}{2^{2n-4}}\sum_{k_1,k_2} \rho_{k_1,k_2,i}^{11},
\end{equation}
where $k_1$ and $k_2$ satisfy definition \ref{def3}.

We can prove that the trace distance between any two of the four states is $O(\frac{1}{2^n})$.
\begin{enumerate}
\renewcommand{\labelenumi}{(\theenumi)}
\item The trace distance between $\rho_{odd}^{00}$ and $\rho_{odd}^{11}$ is $D(\rho_{odd}^{00},\rho_{odd}^{11})=\frac{1}{2^{n-2}}$ (see Appendix A).
\item The trace distance between $\rho_{odd}^{00}$ and $\rho_{odd}^{01}$ is $D(\rho_{odd}^{00},\rho_{odd}^{01})=\frac{1}{2^{n-2}}$ (see Appendix B).
\item The trace distance between $\rho_{odd}^{00}$ and $\rho_{odd}^{10}$ is $D(\rho_{odd}^{00},\rho_{odd}^{10})=\frac{1}{2^{n-2}}$. The proof is similar as that of $D(\rho_{odd}^{00},\rho_{odd}^{01})$.
\item  By the triangle inequality of the trace distance \cite{11}, we have
\begin{equation}
D(\rho_{odd}^{10},\rho_{odd}^{01})\le D(\rho_{odd}^{00},\rho_{odd}^{10})+D(\rho_{odd}^{00},\rho_{odd}^{01})=\frac{1}{2^{n-3}},
\end{equation}
\begin{equation}
D(\rho_{odd}^{10},\rho_{odd}^{11})\le D(\rho_{odd}^{00},\rho_{odd}^{10})+D(\rho_{odd}^{00},\rho_{odd}^{11})=\frac{1}{2^{n-3}},
\end{equation}
\begin{equation}
D(\rho_{odd}^{01},\rho_{odd}^{11})\le D(\rho_{odd}^{00},\rho_{odd}^{01})+D(\rho_{odd}^{00},\rho_{odd}^{11})=\frac{1}{2^{n-3}}.
\end{equation}
\end{enumerate}

\begin{figure}[htp!]
\begin{center}
\includegraphics[width=6.2cm]{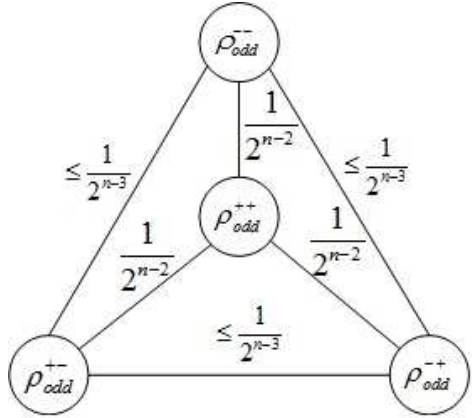}
\end{center}
\caption{\label{fig2}Trace distances between $\rho_{odd}^{00}$ and other three states are $\frac{1}{2^{n-2}}$, so the trace distances between any two states of the other three are no more than $\frac{1}{2^{n-3}}$.}
\end{figure}

As shown in Figure \ref{fig2}, each trace distance between any two of these four states is $O(\frac{1}{2^n})$.

\subsection{Extended QPKE for Two Bits}
\noindent
To extend the QPKE scheme to the two-bit oriented one, two aspects will be modified:
\begin{enumerate}
\renewcommand{\labelenumi}{(\theenumi)}
\item Bob chooses $s\in \{0,1\}^{poly(n)}$ randomly, satisfies $F(s)=(k_1,k_2)$, where $k_1,k_2$ are defined above. Bob generates the $n$-qubit state $\rho_{k_1,k_2,i}^{00}$, and sends $(s,\rho_{k_1,k_2,i}^{00})$ to the public register as his public-key.
\item Alice encrypts $00$ into $\rho_{k_1,k_2,i}^{00}$, $01$ into $\rho_{k_1,k_2,i}^{01}$, $10$ into $\rho_{k_1,k_2,i}^{10}$ and $11$ into $\rho_{k_1,k_2,i}^{11}$, with operations  $I^{\otimes n}$, $I^{\otimes\frac{n}{2}}\otimes Z^{\otimes\frac{n}{2}}$, $Z^{\otimes\frac{n}{2}}\otimes I^{\otimes\frac{n}{2}}$ and $Z^{\otimes n}$ respectively.
\end{enumerate}

Because the trace distance of every pair of the four states is $O(\frac{1}{2^n})$, the QPKE scheme satisfies Eq.(\ref{eqn:one}), so it is a scheme with information theoretic security under CPA.

\subsection{Extended QPKE for Multi Bits}
\noindent
We now extend the QPKE to encrypt $l$ bits. Define $n$-qubit state as:
\begin{equation}
|\Psi_{k_1,k_2,\cdots,k_l,i}^{00\cdots 0}\rangle=\frac{1}{\sqrt{2^l}}\sum_{x_1,\cdots,x_l }|i\oplus x_1k_1 \oplus x_2k_2 \cdots  \oplus x_lk_l\rangle.
\end{equation}
where $k_1,k_2,\cdots,k_l,i \in Z_{2^n}$, $x_1,\cdots,x_l \in \{0,1\}$, each $k_j$ $(j=1,\cdots,l)$ can be divided into l parts $k_j=(k_{j1},\cdots,k_{jl})$, only $W_H(k_{jj})=odd$, and others are even.

We can use
\begin{equation}
\operatorname*{\otimes}\limits_{j=1}^l ((1-x_j)I+x_jZ)^{\otimes \frac{n}{l}}|\Psi_{k_1,k_2,\cdots,k_l,i}^{00\cdots 0}\rangle,
\end{equation}
to represent classical bits $(x_1,\cdots,x_l)$. If we use these $2^l$ states to construct QPKE, the algorithm introduced in Lemma \ref{thm:two} can also be used for decryption. We conjecture that the trace distance of every pair of these mixed states is $O(\frac{1}{2^{n-l}})$, then the extended scheme is also one with information theoretic security under CPA.

\section{Conclusions}
\noindent
We have proposed a definition for the information theoretic security of a quantum public-key encryption scheme, and proved the sufficient condition that a QPKC scheme has information theoretic security if the trace distance between every pair of ciphertext states is less than $1/p(n)$ for every positive polynomial $p(\cdot)$. We present bit-oriented and two-bit-oriented QPKE schemes with a new algorithm structure, and prove that both of them satisfy our security definition. Finally, we extend the QPKE to the multi-bit-oriented case, and conjecture that the scheme is also one with information theoretic security. The information theoretic security of our QPKE schemes depends directly on the new structure of the public-key algorithm that we have introduced here.

\section*{Acknowledgements}
\noindent
This work was supported by the National Natural Science Foundation of China under Grant No. 60573051.

\appendix

\section{}
\noindent
Trace Distance between $\rho_{odd}^{00}$ and $\rho_{odd}^{11}$:
\begin{eqnarray}
\rho_{odd}^{00}-\rho_{odd}^{11}&=&\frac{2}{2^{2n-4}\cdot 2^n \cdot 4}\sum_{k_1,k_2,i}(E_{i,i\oplus k_1}+E_{i,i\oplus k_2}+E_{i\oplus k_1,i}+E_{i\oplus k_1,i\oplus k_1\oplus k_2}\nonumber\\
&+&E_{i\oplus k_2,i}+E_{i\oplus k_2,i\oplus k_1\oplus k_2}+E_{i\oplus k_1\oplus k_2,i\oplus k_1}+E_{i\oplus k_1\oplus k_2,i\oplus k_2}),
\end{eqnarray}
since
\begin{eqnarray}
\sum_{i}E_{i,i\oplus k_1}&=&\sum_{i}E_{i\oplus k_1,i}=\sum_{i}E_{i\oplus k_2,i\oplus k_1\oplus k_2}=\sum_{i}E_{i\oplus k_1\oplus k_2,i\oplus k_2},
\end{eqnarray}
and
\begin{eqnarray}
\sum_{i}E_{i,i\oplus k_2}&=&\sum_{i}E_{i\oplus k_1,i\oplus k_1\oplus k_2}=\sum_{i}E_{i\oplus k_2,i}=\sum_{i}E_{i\oplus k_1\oplus k_2,i\oplus k_1},
\end{eqnarray}
then
\begin{eqnarray}
\rho_{odd}^{00}-\rho_{odd}^{11}&=&\frac{8}{2^{2n-4}\cdot 2^n \cdot 4}\sum_{k_1,k_2,i}(E_{i,i\oplus k_1}+E_{i,i\oplus k_2})\nonumber\\
&=&\frac{8\cdot 2^{n-2}}{2^{2n-4}\cdot 2^n\cdot 4} \sum_{i}(\sum_{k_1}E_{i,i\oplus k_1}+\sum_{k_2}E_{i,i\oplus k_2}) \nonumber\\
&=&\frac{1}{2^{n-3}\cdot 2^n}\sum_{i}\sum_{j \in \Omega_n}E_{i,i\oplus j}\nonumber\\
&=&\frac{1}{2^{n-3}\cdot 2^n}A_{odd}.
\end{eqnarray}

where the $A_{odd}$ is the same as in  (\ref{eqn:Aodd}),
according to (\ref{eqn:trAodd}) we have
\begin{eqnarray}
D(\rho_{odd}^{00},\rho_{odd}^{11})=\frac{1}{2} tr|\rho_{odd}^{00}-\rho_{odd}^{11}|=\frac{1}{2^{n-2}\cdot 2^n}tr|A_{odd}|=\frac{1}{2^{n-2}}.
\end{eqnarray}

\section{}
\noindent
Trace Distance between $\rho_{odd}^{00}$ and $\rho_{odd}^{01}$:
\begin{eqnarray}
&& \rho_{odd}^{00}-\rho_{odd}^{01}=\frac{2}{2^{2n-4}\cdot 2^n \cdot 4}\sum_{k_1,k_2,i}(E_{i,i\oplus k_2}+E_{i,i\oplus k_1\oplus k_2}+E_{i\oplus k_1,i\oplus k_2}+\nonumber\\
&+&E_{i\oplus k_1,i\oplus k_1\oplus k_2}+E_{i\oplus k_2,i}+E_{i\oplus k_2,i\oplus k_1}+E_{i\oplus k_1\oplus k_2,i}+E_{i\oplus k_1\oplus k_2,i\oplus k_1}),
\end{eqnarray}
since
\begin{eqnarray}
\sum_{i}E_{i,i\oplus k_2}&=&\sum_{i}E_{i\oplus k_1,i\oplus k_1\oplus k_2}=\sum_{i}E_{i\oplus k_2,i}=\sum_{i}E_{i\oplus k_1\oplus k_2,i\oplus k_1},
\end{eqnarray}
and
\begin{eqnarray}
\sum_{i}E_{i,i\oplus k_1\oplus k_2}&=&\sum_{i}E_{i\oplus k_1,i\oplus k_2}=\sum_{i}E_{i\oplus k_2,i\oplus k_1}=\sum_{i}E_{i\oplus k_1\oplus k_2,i},
\end{eqnarray}
then
\begin{eqnarray}
&&\rho_{odd}^{00}-\rho_{odd}^{01}\nonumber\\
&=&\frac{8}{2^{2n-4}\cdot 2^n \cdot 4}\sum_{k_1,k_2,i}(E_{i,i\oplus k_2}+E_{i,i\oplus k_1\oplus k_2}) \\
&=&\frac{8}{2^{3n-2}}\sum_{k_{11},k_{12},k_{21}, \atop k_{22},i_1,i_2}(E_{i_1,i_1\oplus k_{21}}\otimes E_{i_2,i_2\oplus k_{22}}
+E_{i_1,i_1\oplus k_{11}\oplus k_{21}}\otimes E_{i_2,i_2\oplus k_{12}\oplus k_{22}}).\nonumber
\end{eqnarray}

Since $W_H(k_{11})=odd$, $W_H(k_{12})=even$, $W_H(k_{21})=even$ and $W_H(k_{22})=odd$, we have
\begin{equation}
\sum_{k_{11}}E_{i_1,i_1\oplus k_{11}\oplus k_{21}}=\sum_{k_{11}}E_{i_1,i_1\oplus k_{11}},
\end{equation}

\begin{equation}
\sum_{k_{22}}E_{i_2,i_2\oplus k_{12}\oplus k_{22}}=\sum_{k_{22}}E_{i_2,i_2\oplus k_{22}}.
\end{equation}

then
\begin{eqnarray}
\rho_{odd}^{00}-\rho_{odd}^{01}&=&\frac{8\cdot 2^{n-2}} {2^{3n-2}}\sum_{i_1,i_2}(\sum_{k_{21},k_{22}}E_{i_1,i_1\oplus k_{21}}\otimes E_{i_2,i_2\oplus k_{22}}+\sum_{k_{11},k_{22}}E_{i_1,i_1\oplus k_{11}}\otimes E_{i_2,i_2\oplus k_{22}})\nonumber\\
&=&\frac{1}{2^{2n-3}}(\sum_{i_1,k_{21}}E_{i_1,i_1\oplus k_{21}}+\sum_{i_1,k_{11}}E_{i_1,i_1\oplus k_{11}})\otimes(\sum_{i_2,k_{22}}E_{i_2,i_2\oplus k_{22}})\nonumber\\
&=&\frac{1}{2^{2n-3}}A_{\frac{n}{2}}\otimes A_{odd,\frac{n}{2}},
\end{eqnarray}
where $A_{odd,\frac{n}{2}}$ is a $2^{\frac{n}{2}}\times 2^{\frac{n}{2}}$ matrix similar to $A_{odd}$, and
\begin{equation*}
A_{\frac{n}{2}}=\left[\begin{array}{cc}
1&1\\
1&1
\end{array}\right]^{\otimes\frac{n}{2}},
\end{equation*}
then we have
\begin{equation*}
tr|A_{\frac{n}{2}}|=2^{\frac{n}{2}},
\end{equation*}
\begin{equation*}
tr|A_{odd,\frac{n}{2}}|=2^{\frac{n}{2}},
\end{equation*}
\begin{eqnarray*}
D(\rho_{odd}^{00},\rho_{odd}^{01})=\frac{1}{2} tr|\rho_{odd}^{00}-\rho_{odd}^{01}|
=\frac{1}{2^{2n-2}}tr|A_{\frac{n}{2}}|\times tr|A_{odd,\frac{n}{2}}|
=\frac{1}{2^{n-2}}.
\end{eqnarray*}

\end{document}